\documentstyle[aps,preprint]{revtex}

\begin{document}

{\tighten
\preprint{\vbox{
\hbox{UCSD/PTH 96--23}
\hbox{hep-ph/9609404}
}}
\title{Update of Heavy Baryon Mass Predictions}
\author{Elizabeth Jenkins\footnotemark\footnotemark}
\address{Department of Physics, University of California at
San Diego, La Jolla, CA 92093}
\bigskip
\date{September 1996}
\maketitle
\widetext
\begin{abstract}
Predictions of unknown heavy baryon masses based on an
expansion in $1/m_Q$, $1/N_c$ and $SU(3)$ breaking are
updated to take into account a recent measurement of the
$\Sigma_c^*$ mass.  Values are given for the two remaining unknown 
charm baryon masses $\Xi_c^\prime$ and $\Omega_c^*$ and the seven 
unknown bottom baryon masses $\Xi_b$, $\Sigma_b$, $\Xi_b^\prime$,
$\Omega_b$, $\Sigma_b^*$, $\Xi_b^*$ and $\Omega_b^*$. 
\end{abstract}
}

\footnotetext{${}^*$Alfred P. Sloan Fellow.}
\footnotetext{${}^\dagger$National Science Foundation Young Investigator.}

The spectrum of baryons containing a single charm or bottom quark 
recently has been analyzed in an expansion in $1/m_Q$, $1/N_c$ and
$SU(3)$ flavor breaking~\cite{hmasses}.  The analysis of Ref.~\cite{hmasses}  
yields a hierarchy of mass relations among heavy quark baryons,
as well as additional relations between heavy baryon
mass splittings and mass splittings of the octet and decuplet baryons.  
From this analysis, it is possible to predict the unknown charm and
bottom baryon masses to varying accuracies, where the errors on the predicted
values reflect both experimental errors of measured baryon masses and the
expected theoretical accuracies of the mass relations.  

Recently, a very precise measurement of the $\Sigma_c^*$ mass has been 
reported by CLEO~\cite{cleonew}.  The most precise flavor-$27$
mass relations~\cite{savage,hmasses} of the charm baryons determine
$\Xi_c^\prime$ and $(\Sigma_c^* + \Omega_c^*)$,
but not $\Sigma_c^*$ and $\Omega_c^*$ separately.  Given the $\Sigma_c^*$ 
mass, it is possible to pin down the $\Omega_c^*$ mass.  
With the entire charm baryon spectrum
determined rather precisely, prediction of the bottom baryon spectrum from 
the charm baryon spectrum is improved.  The mass predictions of 
Ref.~\cite{hmasses} are updated in this note.  Mass predictions for 
individual bottom baryon masses are given.                          
   
The presently measured charm baryon masses are 
\begin{eqnarray}
&&\Lambda_c = 2285.0 \pm 0.6~{\rm MeV} \cite{pdg}, \nonumber\\
&&\Xi_c = 2467.7 \pm 1.2~{\rm MeV}\cite{pdg}, \nonumber\\
&&\Sigma_c = 2452.9 \pm 0.6~{\rm MeV}\cite{pdg,e687,cleoi}, \\
&&\Omega_c^0 = 2699.9 \pm 2.9~{\rm MeV}\cite{e687i}, \nonumber\\
&&\Sigma_c^* = 2518.6 \pm 2.2~{\rm MeV}\cite{cleonew}, \nonumber\\
&&\Xi_c^* = 2644.0 \pm 1.6~{\rm MeV}\cite{cleoii}. \nonumber
\end{eqnarray}
An observation of the $\Xi_c^\prime \sim 2560~{\rm MeV}$ with an
error bar of order $15$~MeV has been reported by the 
WA89 Collaboration~\cite{wa89}.  The $\Omega_c^*$ has never been observed.
At present, only the bottom baryon mass 
\begin{equation}
\Lambda_b = 5623 \pm 5 \pm 4~{\rm MeV} \cite{cdf}
\end{equation}
is accurately measured.  Reported measurements of $\Sigma_b^{(*)}$ by 
DELPHI~\cite{delphi} are not used.   

There are two very accurate mass relations among the charm baryon masses
given in Ref.~\cite{hmasses}:
\begin{equation}\label{mreli}
\frac 1 4 \left[ \left( \Sigma_Q^* - \Sigma_Q \right)
- 2 \left( \Xi^*_Q - \Xi^\prime_Q \right)
+ \left( \Omega^*_Q - \Omega_Q \right) \right]=0,
\end{equation}
with an estimated theoretical error of $0.23$~MeV for $Q=c$ and $0.07$~MeV
for $Q=b$, and
\begin{eqnarray}\label{mrelii}
&&\frac 1 6 \left[ \left( \Sigma_Q + 2 \Sigma_Q^* \right) - 2
\left(\Xi_Q^\prime + 2 \Xi_Q^* \right) + \left( \Omega_Q + 2 \Omega_Q^* \right)
\right]= \nonumber\\
&&\qquad\qquad 
\frac 1 3 \left[\frac 1 4 \left( 2 N - \Sigma - 3 \Lambda + 2 \Xi \right) +
\frac 1 7 \left( 4 \Delta - 5 \Sigma^* - 2 \Xi^* + 3 \Omega \right) \right],
\end{eqnarray}     
with an estimated theoretical accuracy of $1.5$~MeV for $Q=c,b$. 
The linear combination of octet and decuplet masses on the right-hand side of 
Eq.~(\ref{mrelii}) equals $-4.43$~MeV with negligible error.  
Using Eqs.~(\ref{mreli}) and~(\ref{mrelii}), it is possible to predict 
the two unknown charm baryon masses 
$\Xi_c^\prime$ and $\Omega_c^*$ in terms of the other measured charm 
baryon masses:
\begin{eqnarray}
&&\Xi_c^\prime= \frac 1 2 \left( \Sigma_c + \Omega_c \right)
+ (4.43 \pm 1.5)\ {\rm MeV}, \nonumber\\
&&\Omega_c^* = \left( 2 \Xi_c^* - \Sigma_c^* \right) +
(-8.86 \pm 3.0) \ {\rm MeV},
\end{eqnarray}
which yields
\begin{eqnarray}\label{xo}
&&\Xi_c^\prime = 2580.8 \pm 2.1 \ {\rm MeV}, \nonumber\\
&&\Omega_c^* = 2760.5 \pm 4.9 \ {\rm MeV}. 
\end{eqnarray}

With the predicted $\Xi_c^\prime$ and $\Omega_c^*$ masses~(\ref{xo}) 
and the measured charm baryon masses, it is possible
to evaluate any charm baryon mass combination.
The chromomagnetic mass splittings are evaluated to be
\begin{eqnarray}
&&\left( \Sigma_c^* - \Sigma_c \right) = 65.7 \pm 2.3~{\rm MeV}, \nonumber\\
&&\left( \Xi_c^* - \Xi_c^\prime \right) = 63.2 \pm 2.6~{\rm MeV}, \\
&&\left( \Omega_c^* - \Omega_c \right) = 60.6 \pm 5.7~{\rm MeV}. \nonumber
\end{eqnarray}
The spin-averaged sextet masses are 
\begin{eqnarray}
&&\frac 1 3 \left( \Sigma_c + 2 \Sigma_c^* \right) = 2496.7 \pm 1.5~{\rm MeV},
\nonumber\\
&&\frac 1 3 \left( \Xi_c^\prime + 2 \Xi_c^* \right) = 2622.9 \pm 1.3~{\rm MeV},
\\
&&\frac 1 3 \left( \Omega_c + 2 \Omega_c^* \right) = 2740.3 \pm 3.4~{\rm MeV},
\nonumber
\end{eqnarray}
while the sextet mass differences are 
\begin{eqnarray}\label{sesr}
&&\frac 1 3 \left( \Xi_c^\prime + 2 \Xi_c^* \right) - \frac 1 3 \left( \Sigma_c
+ 2 \Sigma_c^* \right) = 126.2 \pm 2.0~{\rm MeV}, \nonumber\\
&&\frac 1 3 \left( \Omega_c + 2 \Omega_c^* \right) - \frac 1 3 \left(
\Xi_c^\prime + 2 \Xi_c^* \right) = 117.4 \pm 3.6~{\rm MeV}. 
\end{eqnarray}      
The $J_\ell^2$ hyperfine splittings in each strangeness sector are
\begin{eqnarray}
&&\frac 1 3 \left( \Sigma_c + 2 \Sigma_c^* \right) - \Lambda_c = 211.7 \pm
1.6~{\rm MeV}, \nonumber\\
&&\frac 1 3 \left( \Xi_c^\prime + 2 \Xi_c^* \right)- \Xi_c 
= 155.2 \pm 1.8~{\rm MeV},
\end{eqnarray}
so the difference of these splittings is large,
\begin{eqnarray}
&&\left[\frac 1 3 \left( \Sigma_c + 2 \Sigma_c^* \right) - \Lambda_c\right]
-\left[\frac 1 3 \left( \Xi_c^\prime + 2 \Xi_c^* \right)- \Xi_c\right]
\nonumber\\
&&\qquad\qquad= 56.5 \pm 2.4~{\rm MeV} .
\end{eqnarray}

The bottom baryon masses can be predicted in terms of the charm baryon
masses and the measured $\Lambda_b$ mass.  

The chromomagnetic hyperfine
splittings of the heavy quark baryons are proportional to $1/m_Q$, so
the bottom baryon chromomagnetic splittings can be obtained from the
charm baryon splittings by rescaling by a factor of $\sim m_c/m_b$.
Including renormalization group running, the scale factor is 
$(Z_b/Z_c)(m_c/m_b) \sim 0.24 \pm 0.05$.  Using this scale factor,  
the chromomagnetic mass splittings of the bottom
baryons are predicted to be
\begin{eqnarray}\label{bchrom}
&&\left( \Sigma_b^* - \Sigma_b \right) = 15.8 \pm 3.3~{\rm MeV}, \nonumber\\
&&\left( \Xi_b^* - \Xi_b^\prime \right) = 15.2 \pm 3.2~{\rm MeV}, \\
&&\left( \Omega_b^* - \Omega_b \right) = 14.5 \pm 3.3~{\rm MeV}, \nonumber
\end{eqnarray}
where the errors on the splittings are dominated by the uncertainty of the 
scale factor.  Note that
Eq.~(\ref{mreli}) is essentially an exact relation for the
chromomagnetic splittings.  In addition, by scaling from the charm system, one
concludes that the mass combination
\begin{equation}
\frac 1 6 \left[ 3 \left( \Sigma_b^* - \Sigma_b \right)
- \left( \Xi_b^* - \Xi_b^\prime \right) 
-2 \left( \Omega_b^* - \Omega_b \right)\right] 
\end{equation}
is quite small, and can be at most a few MeV.

The spin-averaged sextet masses of the bottom baryons and the $\Xi_b$
are determined by four mass relations.  There are two very accurate relations,
namely Eq.~(\ref{mrelii})
\begin{equation}\label{mrelbi}
\frac 1 6 \left[ \left( \Sigma_b + 2 \Sigma_b^* \right) - 2
\left(\Xi_b^\prime + 2 \Xi_b^* \right) + \left( \Omega_b + 2 \Omega_b^* \right)
\right] = -4.43 \pm 1.5~{\rm MeV},
\end{equation}
and
\begin{eqnarray}\label{mrelbii}
&&\left\{-\frac 5 8
\left( \Lambda_b - \Xi_b \right)
+ \frac 1 {24} \left[ 3 \left(\Sigma_b + 2
\Sigma_b^*\right) - \left(\Xi^\prime_b + 2 \Xi_b^* \right)
-2 \left(\Omega_b + 2 \Omega_b^*\right)\right]\right\}\nonumber\\
&&\qquad\qquad= \left\{-\frac 5 8
\left( \Lambda_c - \Xi_c \right)
+ \frac 1 {24} \left[ 3 \left(\Sigma_c + 2
\Sigma_c^*\right) - \left(\Xi^\prime_c + 2 \Xi_c^* \right)
-2 \left(\Omega_c + 2 \Omega_c^*\right)\right]\right\}
\pm 1.0~{\rm MeV},   
\end{eqnarray}
where the errors are the estimated theoretical accuracies of the two 
relations.  The charm mass combination on the right-hand side of 
Eq.~(\ref{mrelbii}) equals $37.5 \pm 1.3~{\rm MeV}$, so Eq.~(\ref{mrelbii})
becomes 
\begin{equation}
\left\{-\frac 5 8
\left( \Lambda_b - \Xi_b \right)
+ \frac 1 {24} \left[ 3 \left(\Sigma_b + 2
\Sigma_b^*\right) - \left(\Xi^\prime_b + 2 \Xi_b^* \right)
-2 \left(\Omega_b + 2 \Omega_b^*\right)\right]\right\}
= 37.5 \pm 1.6~{\rm MeV}. \nonumber  
\end{equation} 
There are two additional mass relations which are less accurate:
\begin{equation}
\left( \Lambda_b - \Xi_b \right) = \left( \Lambda_c - \Xi_c \right)
\pm 4.8~{\rm MeV},
\end{equation}
and
\begin{eqnarray}
&&-\frac 1 3 \left( \Lambda_b + 2 \Xi_b \right) 
+ \frac 1 {18} \left[ 3 \left(
\Sigma_b + 2 \Sigma_b^* \right) + 2 \left( \Xi_b^\prime + 2 \Xi_b^* \right) +
\left( \Omega_b + 2 \Omega_b^* \right) \right]\nonumber\\
&&\qquad\qquad =
-\frac 1 3 \left( \Lambda_c + 2 \Xi_c \right) 
+ \frac 1 {18} \left[ 3 \left(
\Sigma_c + 2 \Sigma_c^* \right) + 2 \left( \Xi_c^\prime + 2 \Xi_c^* \right) +
\left( \Omega_c + 2 \Omega_c^* \right) \right]
\pm 5.1~{\rm MeV}.
\end{eqnarray}
Since 
$(\Lambda_c - \Xi_c)= -182.7 \pm 1.3~{\rm MeV}$ experimentally, the first
equation becomes
\begin{equation}\label{mrelbiv}
\left( \Lambda_b - \Xi_b \right) = -182.7 \pm 5.0~{\rm MeV}.
\end{equation}
The charm baryon mass combination in the second equation is evaluated to be
$172.6 \pm 1.3$~MeV, so 
\begin{equation}\label{mrelbv}
-\frac 1 3 \left( \Lambda_b + 2 \Xi_b \right) 
+ \frac 1 {18} \left[ 3 \left(
\Sigma_b + 2 \Sigma_b^* \right) + 2 \left( \Xi_b^\prime + 2 \Xi_b^* \right) +
\left( \Omega_b + 2 \Omega_b^* \right) \right] =
172.6 \pm 5.3~{\rm MeV}.
\end{equation}
Combined with the measured $\Lambda_b$ mass,  
$\Lambda_b = 5623.0 \pm 6.4~{\rm MeV}$, Eq.~(\ref{mrelbiv}) implies
\begin{equation}
\Xi_b = 5805.7 \pm 8.1~{\rm MeV}.
\end{equation}
The spin-averaged mass of the bottom antitriplet is evaluated to be
\begin{equation}
\frac 1 3 \left( \Lambda_b + 2 \Xi_b \right) = 5744.8 \pm 5.8~{\rm MeV}.
\end{equation}
Eliminating the $\Lambda_b$ and $\Xi_b$ masses from the remaining relations 
yields three mass relations involving the three spin-averaged sextet masses 
of the bottom baryons,
\begin{eqnarray}
&&\frac 1 6 \left[ \left( \Sigma_b + 2 \Sigma_b^* \right) - 2
\left(\Xi_b^\prime + 2 \Xi_b^* \right) + \left( \Omega_b + 2 \Omega_b^* \right)
\right] = -4.43 \pm 1.5~{\rm MeV}, \nonumber\\
&&\frac 1 {24} \left[ 3 \left(\Sigma_b + 2
\Sigma_b^*\right) - \left(\Xi^\prime_b + 2 \Xi_b^* \right)
-2 \left(\Omega_b + 2 \Omega_b^*\right)\right]= -76.7 \pm 3.5~{\rm MeV}, \\
&&\frac 1 {18} \left[ 3 \left(
\Sigma_b + 2 \Sigma_b^* \right) + 2 \left( \Xi_b^\prime + 2 \Xi_b^* \right) +
\left( \Omega_b + 2 \Omega_b^* \right) \right] = 5917.4 \pm 7.9~{\rm MeV}. 
\nonumber 
\end{eqnarray}
The extracted spin-averaged mass combinations are 
\begin{eqnarray}
&&\frac 1 3 \left( \Sigma_b + 2 \Sigma_b^* \right) = 5834.7 \pm 8.7~{\rm
MeV}, \nonumber\\
&&\frac 1 3 \left( \Xi_b^\prime + 2 \Xi_b \right) = 5961.0 \pm 8.2~{\rm MeV},
\\
&&\frac 1 3 \left( \Omega_b + 2 \Omega_b^* \right) = 6078.4 \pm 10.9~{\rm MeV}.
\nonumber
\end{eqnarray}

The precision of the above
extraction of $\Xi_b$ and the spin-averaged sextet masses
presently
is limited by the theoretical accuracies of the two least accurate mass
relations and by the experimental error of the $\Lambda_b$ mass measurement.
Improved accuracy of the $\Lambda_b$ value or accurate measurement of other 
bottom baryon masses will lead to more precise predictions.
Note that there are correlations amongst the spin-averaged mass 
values so that 
certain linear combinations involving
the spin-averaged mass combinations
are known much more accurately than the spin-averaged masses themselves.
For example,  
any linear combination of the two most accurate mass relations 
Eqs.~(\ref{mrelbi}) and~(\ref{mrelbii}) is predicted very accurately.
One such linear combination is
\begin{eqnarray}
&&\left[\frac 1 3 \left( \Sigma_b + 2 \Sigma_b^* \right) - \Lambda_b\right]
-\left[\frac 1 3 \left( \Xi_b^\prime + 2 \Xi_b^* \right)- \Xi_b\right]
\nonumber\\
&&\qquad\qquad= 56.5 \pm 2.8~{\rm MeV} .
\end{eqnarray}

Finally, values for the individual bottom baryon masses masses can be
obtained by combining the extracted chromomagnetic and spin-averaged
masses: 
\begin{eqnarray}
&&\Xi_b = 5805.7 \pm 8.1~{\rm MeV}, \nonumber\\
&&\Sigma_b = 5824.2 \pm 9.0~{\rm MeV}, \nonumber\\
&&\Sigma_b^* = 5840.0 \pm 8.8~{\rm MeV}, \nonumber\\
&&\Xi_b^\prime = 5950.9 \pm 8.5~{\rm MeV}, \\
&&\Xi_b^* = 5966.1 \pm 8.3~{\rm MeV}, \nonumber\\
&&\Omega_b = 6068.7 \pm 11.1~{\rm MeV}, \nonumber\\
&&\Omega_b^* = 6083.2 \pm 11.0~{\rm MeV}. \nonumber
\end{eqnarray}
Again, the values of the individual bottom baryon masses are correlated so 
that many linear combinations are known much more accurately than the 
individual masses themselves.  The uncertainty in the spin-averaged sextet
masses is significantly larger than the uncertainty in the chromomagnetic
sextet splittings, for example.
Improved accuracy of the $\Lambda_b$ measurement
and accurate measurement of other bottom baryon masses in the future will
lead to more precise determinations of the remaining unknown masses.  

\section*{Acknowledgments}

This work was supported in part by the Department of Energy
under grant DOE-FG03-90ER40546.  E.J. also was supported in part by NYI
award PHY-9457911 from the National Science Foundation and by a research
fellowship from the Alfred P. Sloan Foundation.

\end{document}